\documentclass[11pt]{article}

\usepackage[final]{acl}

\usepackage{times}
\usepackage{latexsym}
\usepackage{amsmath}
\usepackage{booktabs}
\usepackage{multirow}
\usepackage{enumitem}
\usepackage{colortbl}
\usepackage[most]{tcolorbox}  
\usepackage{tabularx}
\usepackage[T1]{fontenc}
\usepackage[utf8]{inputenc}

\usepackage{microtype}

\usepackage{inconsolata}

\usepackage{graphicx}

\title{Dynamics of Cognitive Heterogeneity: Investigating Behavioral Biases in Multi-Stage Supply Chains with LLM-Based Simulation 
}

\author{
 \textbf{Jiuyun Jiang\textsuperscript{1,2}},
 \textbf{Yuecheng Hong\textsuperscript{1}},
 \textbf{Bo Yang\textsuperscript{1,2}},
 \textbf{Jin Yang\textsuperscript{1}},
\\
 \textbf{Guangxin Jiang\textsuperscript{1}\thanks{Corresponding author.}},
 \textbf{Xiaomeng Guo\textsuperscript{2}},
 \textbf{Guang Xiao\textsuperscript{2}}
\\
 Harbin Institute of Technology\textsuperscript{1},
 The Hong Kong Polytechnic University\textsuperscript{2}
\\
\texttt{\{jiuyunjiang, 2023112391, boyang, 22B910011\}@stu.hit.edu.cn},\\
\texttt{gxjiang@hit.edu.cn},
\texttt{\{xiaomeng.guo, guang.xiao\}@polyu.edu.hk}
}

\begin{document}
\maketitle

\begin{abstract}

Modeling coordination among generative agents in complex multi-round decision-making presents a core challenge for AI and operations management. Although behavioral experiments have revealed cognitive biases behind supply chain inefficiencies, traditional methods face scalability and control limitations. We introduce a scalable experimental paradigm using Large Language Models (LLMs) to simulate multi-stage supply chain dynamics. Grounded in a Hierarchical Reasoning Framework, this study specifically analyzes the impact of cognitive heterogeneity on agent interactions. Unlike prior homogeneous settings, we employ DeepSeek and GPT agents to systematically vary reasoning sophistication across supply chain tiers. Through rigorously replicated and statistically validated simulations, we investigate how this cognitive diversity influences collective outcomes. Results indicate that agents exhibit myopic and self-interested behaviors that exacerbate systemic inefficiencies. However, we demonstrate that information sharing effectively mitigates these adverse effects. Our findings extend traditional behavioral methods and offer new insights into the dynamics of AI-enabled organizations. This work underscores both the potential and limitations of LLM-based agents as proxies for human decision-making in complex operational environments.

\end{abstract}

\section{Introduction}
\label{sec:introduction}

Human-subject experiments are foundational to behavioral science, revealing how cognitive biases and bounded rationality shape complex decision-making. Classic studies, such as the Beer Distribution Game, have shown how these factors contribute to systemic inefficiencies like the bullwhip effect in supply chains~\cite{sterman1989modeling, croson2006behavioral, croson2013behavioral, steckel2004supply}. These inefficiencies incur significant real-world costs in inventory and revenue. However, traditional experiments face constraints in scalability, cost, and experimental control, which restrict their reproducibility and generalizability~\cite{fehr1999theory}. These challenges necessitate alternative approaches to study human-like strategic behavior at scale.

Recent progress in Large Language Models (LLMs) presents new avenues for behavioral research~\cite{park2023generative, cui2025large}. As computational agents, LLMs enable large-scale, repeatable experiments in which individual characteristics can be precisely defined~\cite{brown2020language, schaeffer2023emergent}. Crucially, recent findings indicate that LLMs can align closely with human judgment in context-sensitive evaluations, suggesting they effectively capture the nuances of human reasoning and inherent biases~\cite{davidson2025multimodal}. While prior studies focus on horizontal social interactions~\cite{park2023generative}, static games~\cite{akata2025playing}, or unstructured operational tasks~\cite{kirshner2024artificial}, they lack the institutional realism required for complex economic environments. In contrast, our framework integrates LLM decision modules with a rule-based architecture to systematically evaluate multi-period behavioral biases within a vertically structured supply chain. Consequently, our approach moves beyond traditional settings to offer a scalable and highly controlled alternative to human laboratory experiments. In addition, many existing studies lack repeated statistical validation, limiting their generalizability to complex, real-world decision-making.

\begin{figure*}[t]
\centering
\includegraphics[width=2\columnwidth]{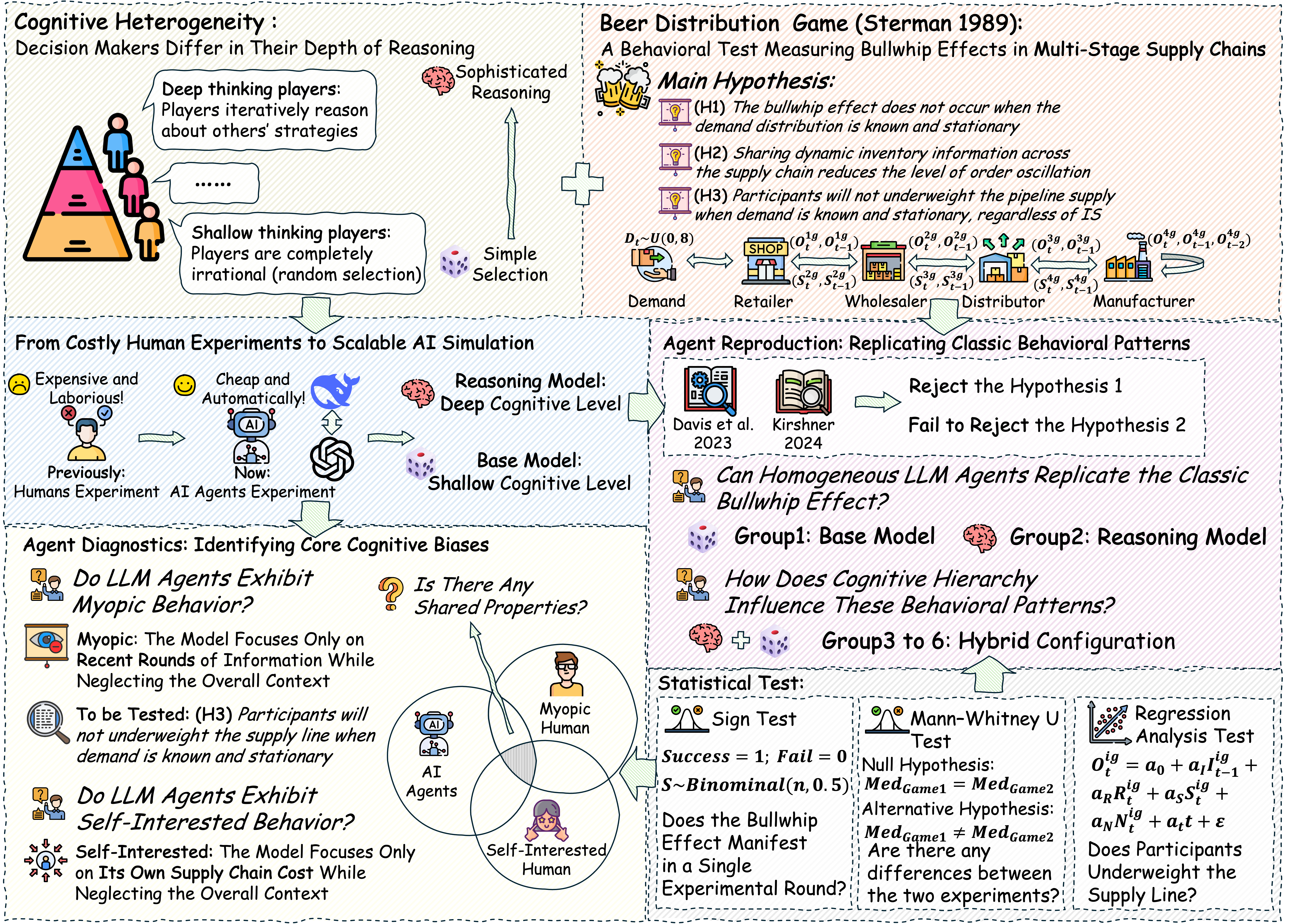} 
\caption{Overview of the experimental workflow and analytical framework.}
\label{fig:our_work}
\end{figure*}

To address these gaps, we deploy LLM agents within the Beer Distribution Game, a dynamic multi-stage supply chain simulation known for its coordination challenges. In this setting, minor demand changes can propagate and amplify, resulting in inventory cycles and performance losses. Leveraging a Hierarchical Reasoning Framework, we vary agent sophistication across tiers to examine how strategic heterogeneity influences system behavior. Unlike prior work, we conduct multi-round replications with rigorous statistical testing to ensure robust and interpretable findings (Figure \ref{fig:our_work}). Our main contributions are summarized as follows:

\begin{itemize}
    \item We present an experimental paradigm using LLM agents that replicates human-like behavioral patterns, including the bullwhip effect, and reveals greater decision stability and statistical clarity compared to human data.
    \item We show that LLM agents exhibit myopic decision-making, which contributes to the bullwhip effect, aligning with observed human behavior.
    \item We find that self-interested behavior among agents reduces overall system performance, and that transparent information sharing significantly improves coordination and mitigates these adverse effects.
\end{itemize}

\section{Related Work}
\label{sec:related_works}

Behavioral operations research has long established that supply chain inefficiencies, such as the bullwhip effect, are deeply rooted in cognitive biases and bounded rationality~\cite{sterman1989modeling, fehr1999theory, croson2006behavioral, croson2013behavioral, steckel2004supply}. To systematically investigate such phenomena, operations and management science rely heavily on a structured methodological paradigm: abstracting complex industrial networks into controlled operational frameworks. This approach isolates specific mechanisms and yields clear theoretical insights, as widely demonstrated in studies on disruption risks, renegotiation dynamics, and technology adoption~\cite{birge2023disruption, katok2025renegotiations, ang2017disruption, dong2023impact}. However, populating these frameworks with traditional human-subject experiments introduces severe constraints in scalability, cost, and experimental control.

LLMs offer a scalable alternative to study cognitive drivers in complex systems~\cite{brown2020language, achiam2023gpt, wei2022emergent, schaeffer2023emergent, bubeck2023sparks, bommasani2021opportunities, chen2021evaluating, park2023generative}. As computational agents, LLMs successfully replicate strategic human behavior in classic games and social dilemmas~\cite{akata2025playing, zhou2022large, webb2023emergent, argyle2023out, kosinski2023theory, lin2021truthfulqa, mei2023turing, wang2025llms}. Building on this capability, recent research has rapidly expanded into organizational simulation and supply chain management. For instance, LLM-based systems are now utilized to interpret optimization outcomes~\cite{li2023large} and facilitate autonomous consensus seeking to mitigate the bullwhip effect~\cite{jannelli2026agentic}. Concurrently, the computational linguistics community has developed extensive frameworks to evaluate collaboration and competition within large-scale multi-agent environments~\cite{wang2025megaagent, zhu2025multiagentbench, agashe2025llm, huang2025licomemory}.

Despite these substantial advancements, two critical gaps remain. First, most studies focus on static or structurally simple settings, leaving the performance of LLM agents in highly dynamic, multi-period environments largely unexplored~\cite{akata2025playing, webb2023emergent, kosinski2023theory}. Second, existing multi-agent simulations frequently deploy homogeneous agents or treat LLMs as uniform behavioral proxies, which obscures the impact of cognitive heterogeneity on group dynamics~\cite{schaeffer2023emergent, jia2024decision, chang2024survey, lin2023toward, quan2024invagent, mannekote2025can, cui2025large}. In real-world organizations, strategic diversity is both prevalent and highly consequential, yet its interaction effects remain insufficiently studied in synthetic settings.

Our work addresses these gaps by embedding a Hierarchical Reasoning Framework within a dynamic, LLM-driven supply chain. By aligning with the classical operations paradigm of structural abstraction, we ensure methodological validity while modeling cognitively heterogeneous agents. This approach moves beyond simple replication to reveal human-like behavioral tendencies and demonstrate how transparent information sharing can mitigate systemic inefficiencies,  providing a foundation for designing resilient socio-technical systems.

\section{Preliminaries}
\label{sec:preliminaries}

\subsection{The Beer Distribution Game Environment}

Our study is based on the canonical Beer Distribution Game (Appendix \ref{Appendix: beer game}), a standard model used to examine inventory management and coordination in multistage supply chains. To ensure direct comparability with prior work involving human participants and artificial agents, we implement the simulation environment strictly following established conventions in the operations management literature (see Appendix \ref{Appendix: experimental design})~\cite{croson2006behavioral, davis2023replication, kirshner2024artificial}.

Each simulated supply chain comprises four agents, representing the retailer (S1), wholesaler (S2), distributor (S3), and manufacturer (S4). Over $T$ discrete time periods, agents interact only with their immediate upstream and downstream neighbors. Together, they aim to manage inventory and fulfill external demand. The system evolves according to deterministic update rules governing shipments $S_t^{i,g}$ and inventory levels $I_t^{i,g}$. Retail demand in period $t$, denoted by $D_t$, is independently drawn from a discrete uniform distribution over the integers from 0 to 8. At each time step, agent $i$ in team $g$ chooses an order quantity $O_t^{i,g}$, which constitutes the agent decision variable. The full set of operational equations is defined as follows:

{\small
\begin{align*}
S_t^{i,g} &= \min\left\{ D_t, \max\left[ I_{t-1}^{i,g} + S_{t-2}^{i+1,g}, 0 \right] \right\} 
\quad i=1, \\
&= \min\left\{ O_{t-2}^{i-1,g}, \max\left[ I_{t-1}^{i,g} + S_{t-2}^{i+1,g}, 0 \right] \right\}
\quad i=2,3, \\
&= \min\left\{ O_{t-2}^{i-1,g}, \max\left[ I_{t-1}^{i,g} + O_{t-3}^{i,g}, 0 \right] \right\}
\quad i=4,
\end{align*}}with corresponding inventory updates, where negative inventory represents a backlog:
$$
I_t^{i,g} =
\begin{cases}
  I_{t-1}^{i,g} + S_{t-2}^{i+1,g} - D_t & \text{for } i=1, \\
  I_{t-1}^{i,g} + S_{t-2}^{i+1,g} - O_{t-2}^{i,g} & \text{for } i=2,3, \\
  I_{t-1}^{i,g} + O_{t-3}^{i,g} - O_{t-2}^{i,g} & \text{for } i=4.
\end{cases}
$$

Each participant is incentivized to minimize their own individual total cost, which is defined as the sum of their holding and backlog costs over all periods. The cost for a participant at stage $i$ is given by:
\begin{equation*}
C_i(T) = \sum_{t=1}^{T} \left[ h^i \max\left\{ I_t^{i}, 0 \right\} - s^i \min\left\{ I_t^{i}, 0 \right\} \right],
\end{equation*}
where $C_i(T)$ is the total cost for the participant at stage $i \in \{1, 2, 3, 4\}$ over $T$ periods, and $h^i$ and $s^i$ are their respective unit holding and backlog costs.

\subsection{Computational Baselines}

Before detailing our generative agent approach, we establish the necessity of utilizing advanced language models by evaluating traditional computational baselines. The heuristic benchmark considered in this study is a demand tracking policy, where each stage adjusts its target inventory level based on recent realized demand instead of maintaining a fixed base stock level. This mechanism aligns replenishment decisions with observed consumption while accounting for lead time and backlog. The order decision at stage $m$ and period $t$ is given by
\begin{equation*}
O_{m,t} = \min\big(\max(0, O^*_{m,t}), c_m\big),
\end{equation*}
where the unconstrained order quantity $O^*_{m,t}$ is computed as
\begin{equation*}
O^*_{m,t}
=
I^*_{m,t}
-
I_{m,t-1}
-
B_{m+1,t-1}
-
\sum_{\Delta t = 1}^{L_m} R_{m,t-\Delta t}.
\end{equation*}
The desired inventory level $I^*_{m,t}$ is defined as
\begin{equation*}
I^*_{m,t}
=
\bar{S}_{m,t-1} L_m
+
B_{m,t-1},
\end{equation*}
where the moving average demand is
\begin{equation*}
\bar{S}_{m,t-1}
=
\frac{1}{L_{\max}}
\sum_{\Delta t = 1}^{L_{\max}} S_{m,t-\Delta t}.
\end{equation*}
Here, $m$ denotes the supply chain stage and $t$ the period. $I_{m,t-1}$ represents on hand inventory, $B_{m,t-1}$ and $B_{m+1,t-1}$ denote backlog, and $R_{m,t-\Delta t}$ captures incoming shipments within lead time $L_m$.

We further consider two reinforcement learning baselines. Independent Proximal Policy Optimization (IPPO) trains each agent independently with shared parameters~\cite{de2020independent}, while Multi Agent Proximal Policy Optimization (MAPPO) incorporates a centralized value function to leverage global information during training~\cite{yu2022surprising}. For both methods, the reward is defined as stage wise profit:
\begin{equation*}
R_{m,t} = p_m \cdot S_{m,t} - r_m \cdot R_{m,t} - k_m \cdot B_{m,t} - h_m \cdot I_{m,t},
\end{equation*}
where $p_m$ denotes the sale price, $r_m$ the order cost, $k_m$ the backlog cost, and $h_m$ the holding cost.

Empirically, a sign test shows the tracking demand policy lacks a statistically significant bullwhip effect ($p = 0.37$). This indicates rigid heuristics fail to capture the systemic inefficiencies characteristic of human bounded rationality. Conversely, while both IPPO and MAPPO produce significant demand amplification ($p < 0.001$) and align with certain behavioral deviations, they require strictly defined state spaces and extensive environment specific training. These constraints severely limit their adaptability to novel operational structures. In contrast, LLM-driven agents inherently process rich contextual information and demonstrate immediate adaptability without domain specific retraining. This makes them a vastly superior and more scalable framework for modeling the cognitive flexibility and strategic nuances of human decision making in complex environments.

\subsection{Hypotheses and Experimental Framework}

Motivated by these operational advantages, this study deploys generative agents to advance prior behavioral research by investigating three central hypotheses:
\begin{description}
\item[\textbf{(H1)}] \emph{The bullwhip effect does not occur when the demand distribution is known and stationary.}
\item[\textbf{(H2)}] \emph{Sharing dynamic inventory information (IS) across the supply chain reduces the level of order oscillation.}
\item[\textbf{(H3)}] \emph{Participants will not underweight the pipeline supply when demand is known and stationary, regardless of IS.}
\end{description}

To systematically test these hypotheses, our research empirically validates a Hierarchical Reasoning Framework. This framework proposes that strategic thinking can be modeled in distinct cognitive layers characterized by increasing complexity. A key methodological challenge is to rigorously isolate cognitive depth from other potential confounding variables. To address this, we develop a dual family approach to represent different levels of reasoning. Specifically, the cognitively shallow tier comprises DeepSeek-V3 and GPT-4.1, while the cognitively deep tier incorporates DeepSeek-R1 and the advanced GPT-5. Across widely used reasoning intensive benchmarks, including \textit{AIME}, \textit{GPQA}, \textit{MMLU}, and \textit{SWE-bench}, DeepSeek-R1 and GPT-5 consistently outperform their respective base counterparts. These benchmarks assess mathematical reasoning, graduate level problem solving, and software engineering reasoning, all of which are closely related to structured decision making tasks. Their superior benchmark performance therefore provides empirical support for our experimental classification. In addition, the distinction between reasoning oriented and base configurations is directly reflected in the simulation outcomes. Across both model families, the reasoning configuration consistently yields lower total cost and lower order variance. Detailed results and discussion are provided in Appendix~\ref{Appendix: cognitive choice}.

This design enables a structured evaluation of the framework core claims. The DeepSeek series shares architectural features, allowing control over model lineage while varying reasoning depth. Prior work validates the performance of DeepSeek-V3~\cite{cui2025large} and the advanced reasoning of DeepSeek-R1~\cite{guo2025deepseek}. To ensure generalizability, we also include the GPT series, with GPT-5 serving as a benchmark for deep reasoning. Results are consistent across sampling temperatures~\cite{cui2025large}, supporting the use of a fixed temperature of 1.

The experiment includes two conditions. In the \textbf{Homogeneous} condition, all agents are from the same cognitive tier: \emph{Original} assigns shallow agents to all stages, while \emph{R-Overall} uses deep agents throughout. In the \textbf{Hierarchical} condition, a single deep agent is placed at one of four positions: Retailer (\emph{R-S1}), Wholesaler (\emph{R-S2}), Distributor (\emph{R-S3}), or Manufacturer (\emph{R-S4}). Each of the six configurations is independently replicated 32 times over 20 periods, consistent with previous studies \cite{davis2023replication, kirshner2024artificial}. All agents use Chain of Thought (CoT) prompting to support structured and transparent decision making.

\begin{figure}[h]
\centering
\includegraphics[width=1\columnwidth]{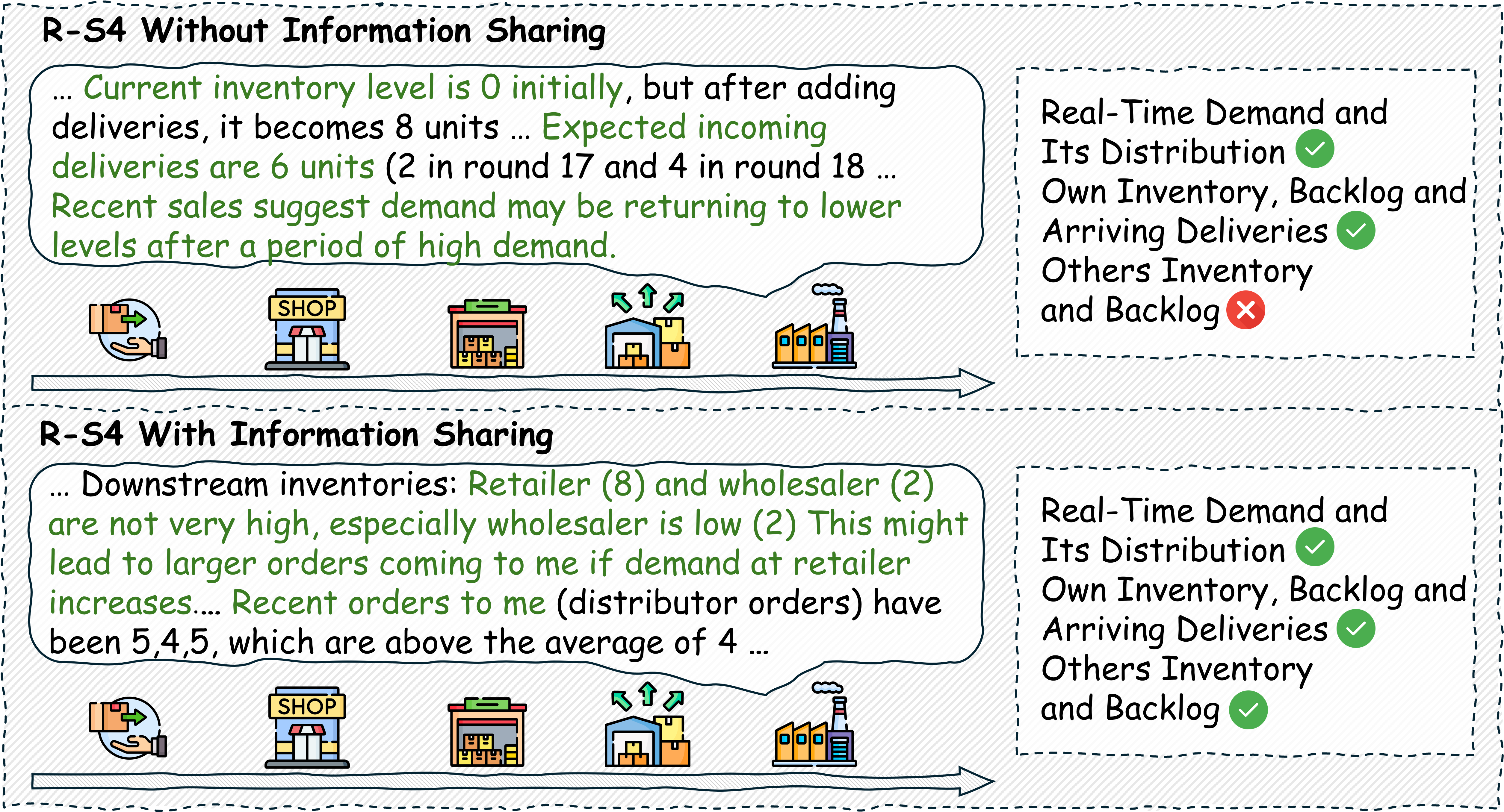} 
\caption{The Mechanism of the Information Structure.}
\label{fig:IS+R}
\end{figure}

To illustrate the core mechanism of the information structure and the agent capability to utilize shared information, we present an example based on the DeepSeek series. Figure~\ref{fig:IS+R} compares the internal reasoning processes of a cognitively deep agent (R-S4) under conditions with and without information sharing. Without information sharing, the agent relies exclusively on local observations. With information sharing, inventory and backlog information from other stages are provided to the language model, thereby enabling more informed decision making. This comparison demonstrates that synthetic agents can effectively leverage information sharing to make more comprehensive decisions. Details of the mechanism design and implementation are provided in Appendix~\ref{Appendix: prompt section}.

\section{Agent Reproduction: Replicating Classic Behavioral Patterns}

To validate our LLM agents as credible proxies for human behavior, we assess their ability to reproduce core behavioral patterns under increasingly complex conditions. We begin by establishing a baseline using homogeneous agent groups. We then introduce cognitive heterogeneity to test whether these patterns persist in more realistic, mixed-agent supply chains.

\subsection{Can Homogeneous LLM Agents Replicate the Classic Bullwhip Effect?}

This section establishes a behavioral baseline by testing whether homogeneous teams of LLM agents can reproduce the classic bullwhip effect. To ensure comparability, both replication studies follow the protocols of prior work~\cite{davis2023replication, kirshner2024artificial}. We evaluate agent performance to test Hypothesis 1 and Hypothesis 2.

\begin{figure}[h]
\centering
\includegraphics[width=1\columnwidth]{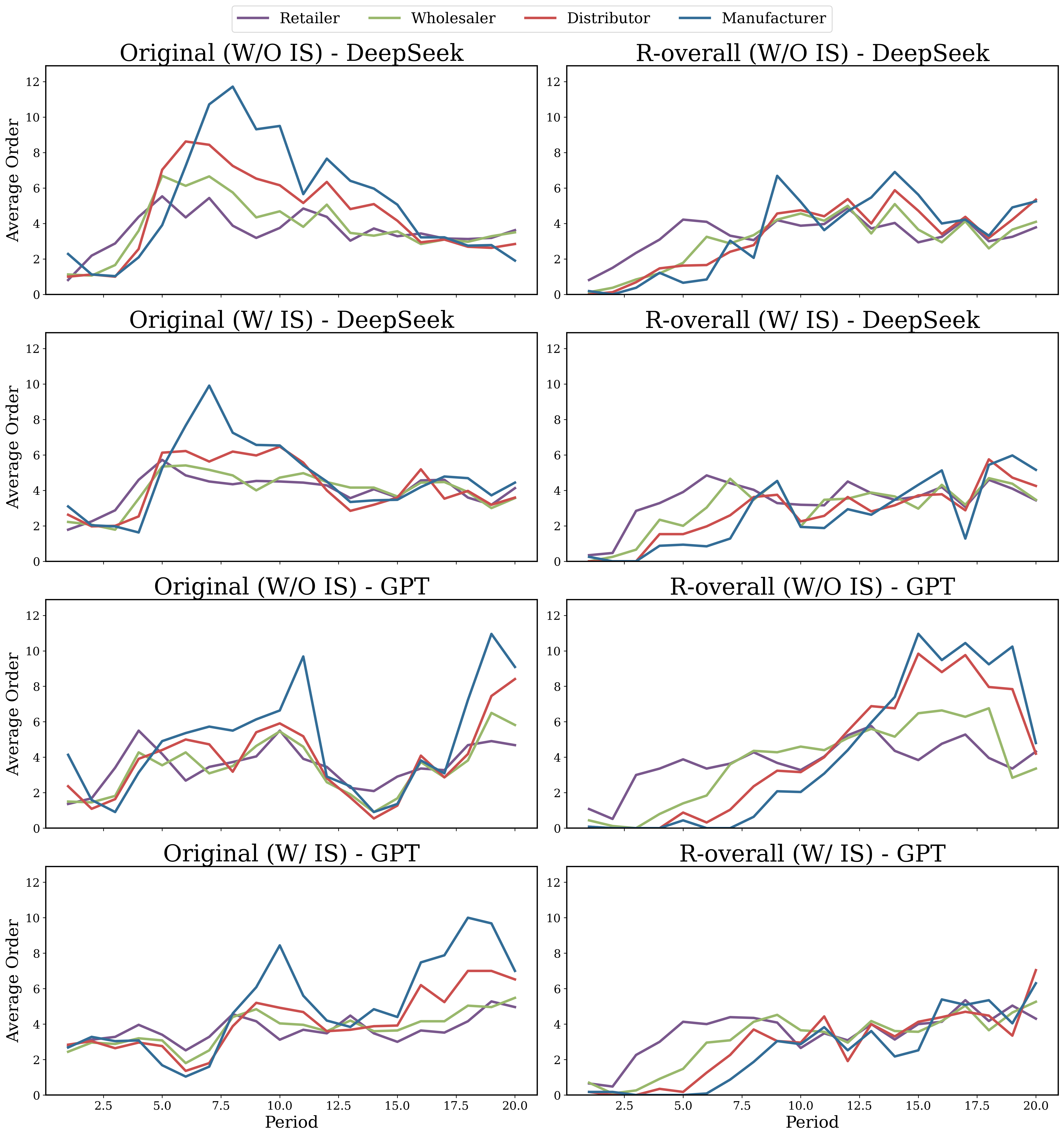} 
\caption{Order amplification: bullwhip effect in homogeneous LLM agent teams (W/ and W/O IS).}
\label{fig:linechart_replication}
\end{figure}

\noindent\textbf{Experimental Results.}
We begin by evaluating Hypothesis~1. Contrary to our initial expectation, a non-parametric sign test ($N=96$, $p < 0.001$) reveals a clear bullwhip effect characterized by significant order amplification along the supply chain (Figure~\ref{fig:linechart_replication}). Quantitatively, order variance increased by $82.3\%$ and $79.8\%$ for the \textit{DeepSeek} Original and R-Overall configurations, respectively. Likewise, the \textit{GPT} configurations exhibited variance increases of $74.2\%$ and $74.3\%$.

\begin{table*}[t] 
\centering
\small
\setlength{\tabcolsep}{12pt} % 调整列间距使表格更美观
\begin{tabular}{l cc cc cc} 
\toprule
& \multicolumn{2}{c}{\textbf{DeepSeek-Original}} 
& \multicolumn{2}{c}{\textbf{DeepSeek-Overall}} 
& \multicolumn{2}{c}{\textbf{UTD (Human)}} \\
\cmidrule(lr){2-3} \cmidrule(lr){4-5} \cmidrule(lr){6-7}
& Mean (S.D.) & Med. 
& Mean (S.D.) & Med. 
& Mean (S.D.) & Med. \\
\midrule
\textbf{W/O IS}  & 39.43 (65.14) & 18.73  & 29.43 (63.54) & 12.29 & 8434 (27620) & 20.46 \\
\textbf{W/ IS}   & 20.15 (29.88) & 11.12  & 17.71 (27.00) & 9.50  & 4347 (21671) & 17.58 \\
\textbf{$p$-value} & 0.001 & \cellcolor{gray!30}$<$0.001 & 0.029 & \cellcolor{gray!30}0.042 & 0.184 & \cellcolor{gray!30}0.221 \\
\textbf{Test}    & $t$-test & \cellcolor{gray!30}M.W. & $t$-test & \cellcolor{gray!30}M.W. & $t$-test & \cellcolor{gray!30}M.W. \\

\midrule \midrule % 双线区分上下两组
& \multicolumn{2}{c}{\textbf{GPT-Original}} 
& \multicolumn{2}{c}{\textbf{GPT-Overall}} 
& \multicolumn{2}{c}{\textbf{UW-Mad. (Human)}} \\
\cmidrule(lr){2-3} \cmidrule(lr){4-5} \cmidrule(lr){6-7}
& Mean (S.D.) & Med. 
& Mean (S.D.) & Med. 
& Mean (S.D.) & Med. \\
\midrule
\textbf{W/O IS}  & 58.45 (89.86) & 23.07 & 31.42 (46.09) & 13.88 & 1148 (6609) & 30.91 \\
\textbf{W/ IS}   & 40.39 (71.44) & 20.70 & 13.72 (9.09)  & 11.88 & 4647 (22490) & 19.83 \\
\textbf{$p$-value} & 0.067 & \cellcolor{gray!30}0.044 & 0.001 & \cellcolor{gray!30}0.018 & 0.116 & \cellcolor{gray!30}0.048 \\
\textbf{Test}    & $t$-test & \cellcolor{gray!30}M.W. & $t$-test & \cellcolor{gray!30}M.W. & $t$-test & \cellcolor{gray!30}M.W. \\
\bottomrule
\end{tabular}

\caption{Comparison of order variance between LLM agent replications and human benchmark studies. S.D. denotes standard deviation, Med. denotes median, and M.W. denotes the Mann--Whitney U test. Human benchmark data are from Davis (2023)~\cite{davis2023replication}. The $p$-values for the M.W. test in the classic human study and a prior GPT-4 replication are 0.028 and 0.573, respectively~\cite{croson2006behavioral}~\cite{kirshner2024artificial}.}

\label{tab:order_variance-merged}
\end{table*}

\begin{figure}[h]
\centering
\includegraphics[width=1\columnwidth]{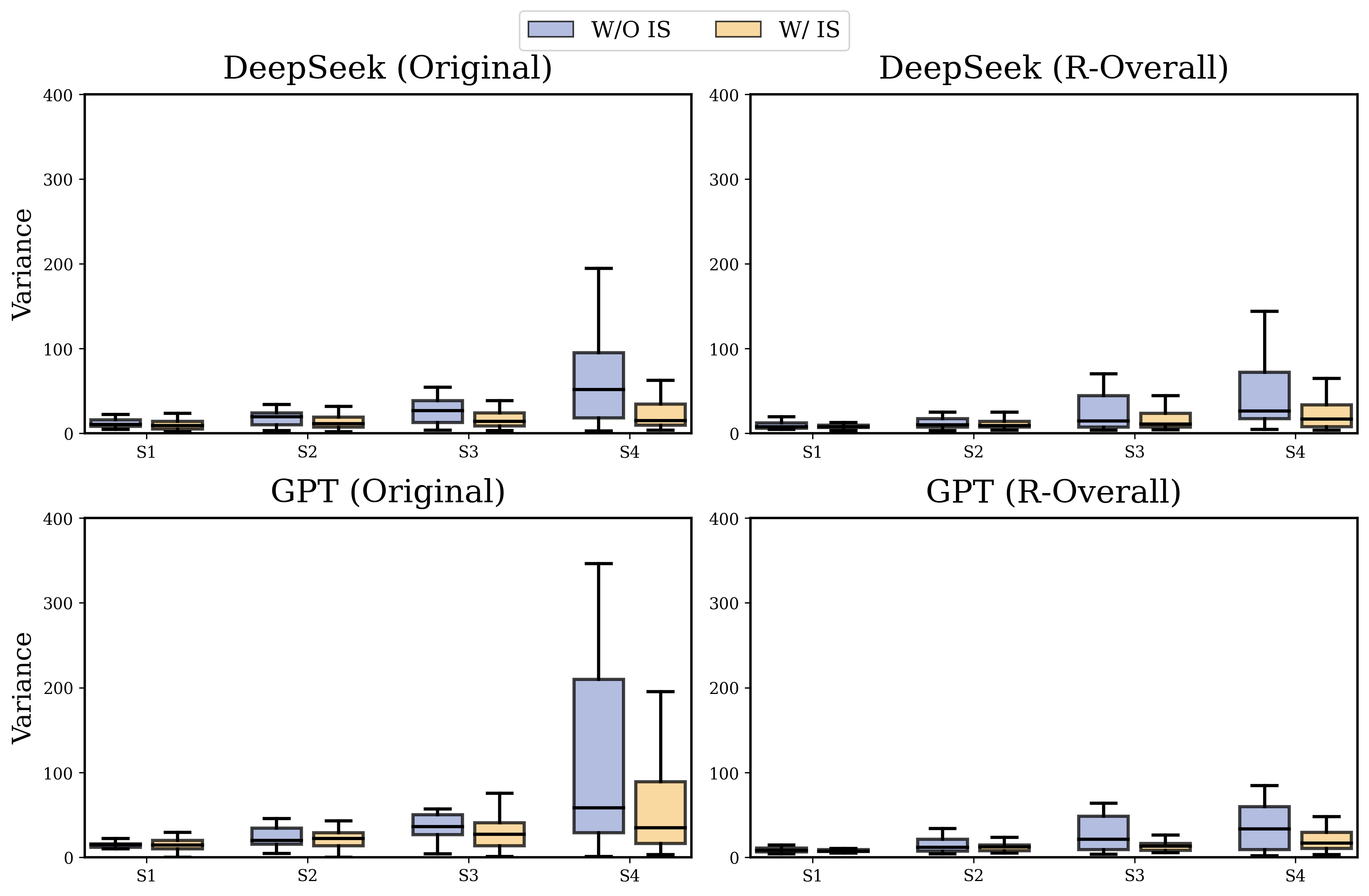} 
\caption{Impact of information sharing on order variance in homogeneous LLM agent teams.}
\label{fig:variance_replication}
\end{figure}

Building on the behavioral baseline, we test Hypothesis~2. This effect has been previously observed in human-subject studies~\cite{croson2006behavioral}. As shown in Figure~\ref{fig:variance_replication} and Table~\ref{tab:order_variance-merged}, IS reduced order variance across all four LLM agent configurations. The Mann-Whitney U test confirmed statistical significance for DeepSeek-Original ($p < 0.001$), DeepSeek-Overall ($p = 0.042$), GPT-Original ($p = 0.044$), and GPT-Overall ($p = 0.018$). In human benchmark data, the effect was significant in the UW-Madison study ($p = 0.048$), but not in the UTD study ($p = 0.221$).

\noindent\textbf{Discussion.}
Our results refute Hypothesis~1. Mirroring human participants in prior studies, LLM agents consistently reproduced the bullwhip effect under idealized and transparent demand conditions. This replication confirms the internal validity of our agent-based framework and highlights its relevance for modeling complex supply chain dynamics. It also demonstrates the potential of LLM-driven simulations to contribute to the field of behavioral operations research.
% Our findings first confirm the rejection of Hypothesis~1. The LLM agents, akin to human participants in canonical studies, consistently generated the bullwhip effect despite stationary and fully transparent demand. This successful replication of a foundational behavioral bias is crucial; it establishes the validity of our agent-based framework to model and investigate complex supply chain phenomena, confirming its ecological relevance for management research.

The test of Hypothesis~2 offers more nuanced insights. While human studies report mixed effects of information sharing~\cite{davis2023replication, croson2006behavioral}, all LLM agent groups showed significant reductions in order variance. This suggests more consistent use of shared information. Agents also showed greater operational stability, with order variances and standard deviations much lower than those observed in human data. Their decisions, though strategically aligned with human behavior, were notably less erratic.

A key difference lies in the statistical profiles. The Mann-Whitney U test confirmed the mitigation effect for most agent groups. This suggests that LLM agents may offer a sensitivity advantage, revealing clearer signals of treatment effects. Our framework thus serves not only as a replication tool but also as a complementary method for theory testing when human data are inconclusive. Further work is needed to understand the implications of this enhanced sensitivity.

\begin{center}
\fbox{
    \parbox{0.94\linewidth}{
        \textbf{Insights:} LLM agents consistently reproduce the bullwhip effect while exhibiting much lower order variance and greater stability than human benchmarks. Their response to information sharing is more uniform and statistically robust. These results demonstrate the framework's enhanced sensitivity for evaluating interventions.
    }
}
\end{center}

\subsection{How Does Cognitive Heterogeneity Influence Behavioral Patterns?}

We subsequently extend our analysis to heterogeneous supply chains, where agents differ in cognitive capabilities. This setting serves a critical validation purpose: real-world organizations are rarely composed of uniform decision-makers. Therefore, it is essential to determine whether the established behavioral paradigms are robustly replicated in mixed-agent populations, or if cognitive heterogeneity alters these fundamental dynamics.

\begin{figure*}[t]
\centering
\includegraphics[width=2\columnwidth]{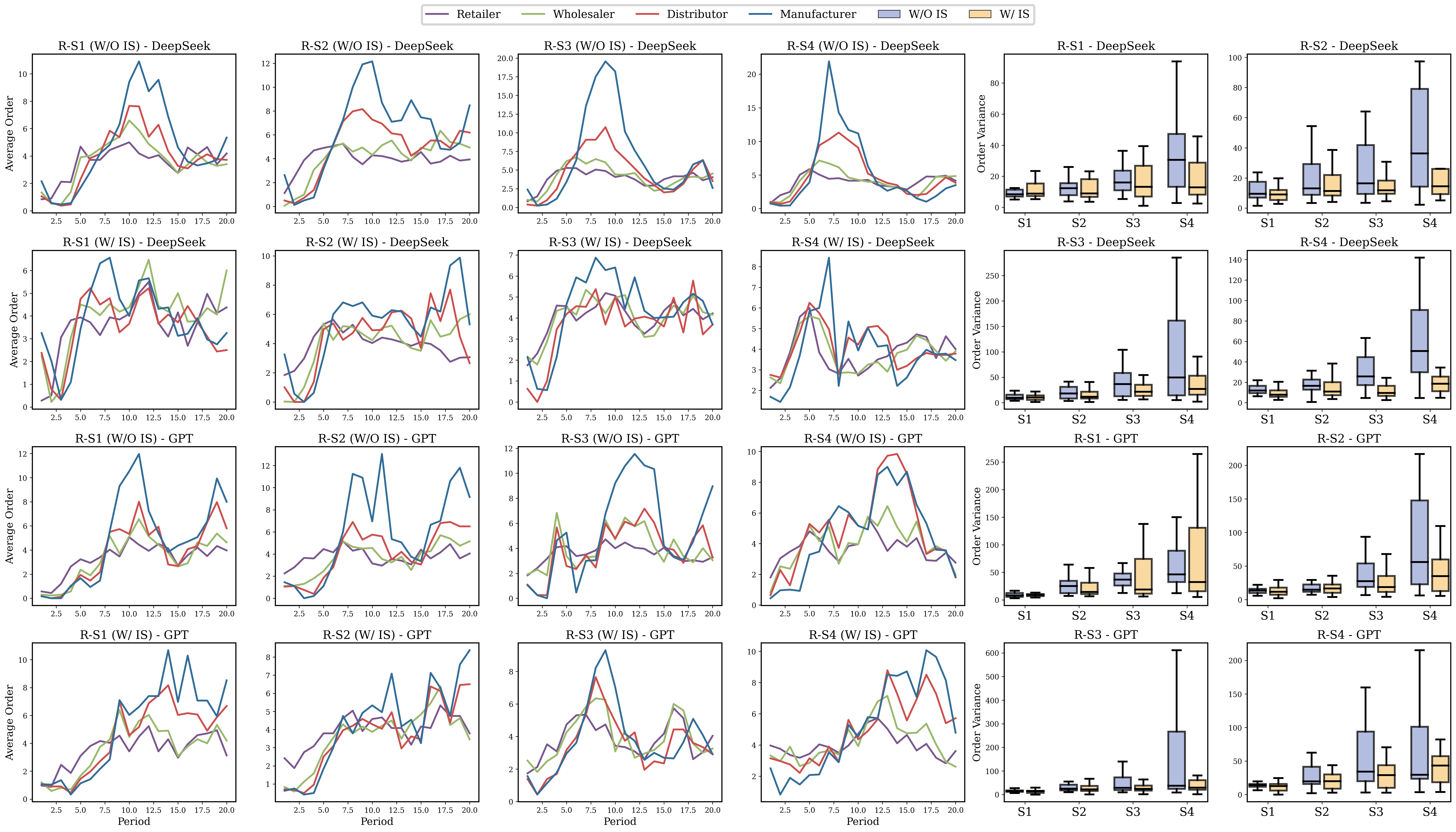} 
\caption{Ordering dynamics and variance in cognitive heterogeneity groups. The results demonstrate the robustness of the bullwhip effect (W/O IS) and the universal efficacy of information sharing (W/ IS) across different heterogeneous configurations. While individual agent sophistication varies, the systemic behavioral patterns remain consistent with the classic Beer Distribution Game paradigm.}
\label{fig:mixed_4x6}
\end{figure*}

\noindent\textbf{Experimental Results.}
Figure~\ref{fig:mixed_4x6} visualizes the order dynamics and variance across distinct cognitive heterogeneity configurations. The top row (W/O IS) confirms that the bullwhip effect persists even in the presence of sophisticated agents. For both DeepSeek and GPT, order oscillations amplify consistently from the retailer (S1) to the manufacturer (S4). This indicates that integrating a single high-reasoning agent is insufficient to eliminate systemic coordination failures driven by local information processing. This pattern is statistically validated by sign tests, showing significant amplification across all heterogeneous configurations ($N = 96$, $p < 0.001$).

We evaluate the effect of interventions (Hypothesis 2). The bottom row (W/ IS) shows that information sharing consistently reduces order fluctuations across agent team compositions. Box plots indicate that order variance with IS (orange) is lower than without IS (blue), supported by statistical results in Table~\ref{tab:pvalues_groups}. Although model-specific differences exist, where DeepSeek shows stronger effects upstream and GPT downstream, the overall pattern is consistent. Information sharing significantly reduces variance across most positions for both models ($p < 0.05$), confirming that the uncertainty-reducing effect of information sharing extends to cognitively diverse AI agents and mirrors human supply chain dynamics.

\begin{table}[h]
\centering
\small
\setlength{\tabcolsep}{3pt}
\begin{tabular}{lcccc}
\toprule
\textbf{Model} & \textbf{R-S1} & \textbf{R-S2} & \textbf{R-S3} & \textbf{R-S4} \\
\midrule
\textbf{DeepSeek} & 0.030 & 0.011 & 0.039 & $<$0.001 \\
\textbf{GPT}  & $<$0.001 & 0.028 & 0.042 & 0.046 \\
\bottomrule
\end{tabular}
\caption{$p$-values for the mitigating effect of information sharing (IS) on order variance across cognitive heterogeneity groups.}
\label{tab:pvalues_groups}
\end{table}

\noindent\textbf{Discussion.}
These findings provide strong evidence that LLM-based agents, even in heterogeneous teams, can replicate key human behavioral patterns. The persistence of the bullwhip effect in mixed groups suggests that systemic structure and information asymmetry often outweigh individual cognitive ability. Similar to human experiments where skilled individuals cannot stabilize a blind supply chain, a single advanced LLM agent cannot prevent demand distortion without structural support such as information sharing. Capturing this dominance of system-level dynamics over individual capability supports the use of LLM agents as valid proxies for complex, multi-round human decision-making. The results indicate that cognitive heterogeneity enhances realism without violating core behavioral patterns in operations research.

\begin{center}
\fbox{
\parbox{0.94\linewidth}{
\textbf{Insights:}
Cognitive heterogeneity does not eliminate systemic behavioral biases. The replication of the bullwhip effect and the efficacy of information sharing in mixed-agent populations confirm that LLMs robustly capture the structural and behavioral dynamics of human supply chains. This validates the use of heterogeneous LLM agents as realistic proxies for studying organizational coordination.
}
}
\end{center}

\section{Agent Diagnostics: Identifying Core Cognitive Biases}

Having established the macro-level effects of heterogeneity and information sharing, we now examine the micro-level cognitive mechanisms. Understanding how LLM agents behave in complex decision-making is essential for their effective use in multi-agent systems. Our experiments reveal a counterintuitive finding: under information asymmetry, improving a single agent’s cognitive ability can worsen overall system performance. This highlights the importance of systemic structures over individual capabilities. We explore this paradox by identifying two core human-like biases.

\begin{figure}[h]
\centering
\includegraphics[width=1\columnwidth]{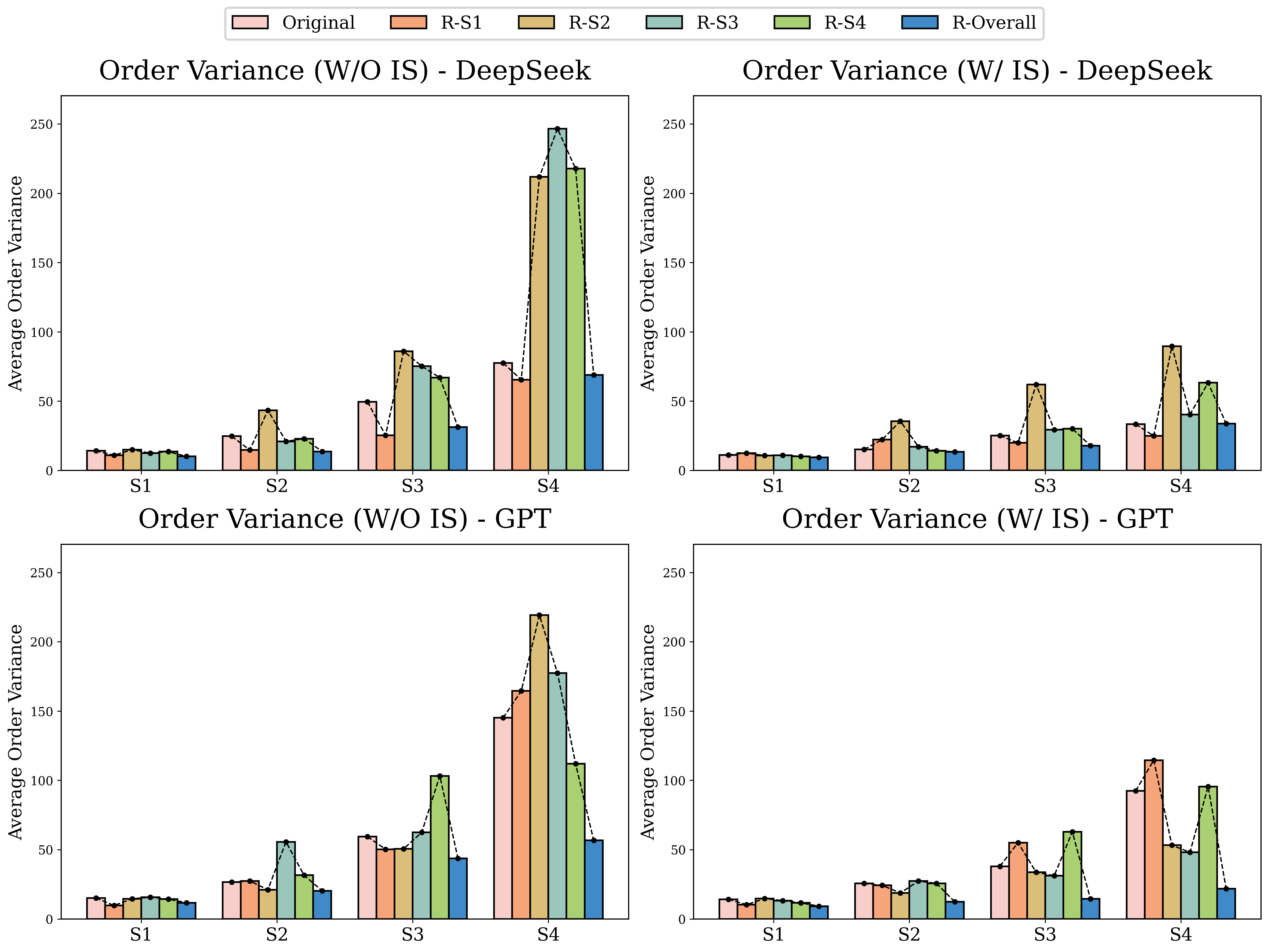} 
\caption{Order variance across supply chain stages for different agent configurations, comparing scenarios without (left) and with (right) information sharing.}
\label{fig:stage_order}
\end{figure}

The left panels of Figure~\ref{fig:stage_order} show that without information sharing, the bullwhip effect persists across models. Enhancing a single agent in upstream positions (R-S2, R-S3, R-S4) increases order variance compared to the homogeneous baseline. GPT tends to produce higher variance than DeepSeek. Two exceptions emerge: full enhancement (R-Overall) leads to clear improvement, and the R-S1 setting shows moderate gains. For GPT, early-stage performance matches the baseline and helps prevent extreme variance at the final stage. DeepSeek in R-S1 yields modest but consistent improvements throughout.

The right panels show that information sharing reduces variance in all configurations. This suggests that enhancing a single agent in an opaque environment may be ineffective, while system-wide transparency offers a reliable solution to mitigate instability driven by myopic behavior.

\subsection{Do LLM Agents Exhibit Myopic Behavior?}

We attribute this performance degradation primarily to human-like myopia, where LLM agents systematically underweight the supply line (inventory in transit). We formalize and test this tendency through controlled experiments as Hypothesis 3.

\noindent\textbf{Experimental Results.}
We evaluate agent ordering behavior using a regression model adapted from Sterman~\cite{sterman1989modeling}:

{\small
\begin{equation*}
\label{eq1}
O_t^{ig} = a_0 + a_I I_{t-1}^{ig} + a_R R_t^{ig} + a_S S_t^{ig} + a_N N_t^{ig} + a_t t + \varepsilon,
\end{equation*}}
where $I_{t-1}^{ig}$ is prior inventory, $R_t^{ig}$ is downstream orders, $S_t^{ig}$ is shipments received from upstream suppliers, $N_t^{ig}$ is total outstanding orders, and $t$ controls for time trends. The coefficient $a_N$ is central for assessing whether agents account for the supply line.

Rational supply line accounting suggests that the coefficients for net stock ($a_N$) and inventory ($a_I$) should be equal, while myopic behavior predicts $a_N > a_I$. As shown in Appendix~\ref{Appendix: regression}, our results consistently find $a_N > a_I$ across all conditions ($p < 0.001$, sign test, $N = 128$), replicating patterns observed in human studies~\cite{croson2006behavioral}. This confirms supply line underweighting as a stable tendency in LLM agents.

\noindent\textbf{Discussion.}
The regression results show that LLM agents exhibit a human-like myopic bias, directly contradicting Hypothesis~3 and rejecting the assumption that agents properly account for pipeline inventory. When a more capable agent is placed upstream, it continues to rely on distorted local information, similar to simpler agents. This persistent myopia leads the agent to misapply its enhanced reasoning, amplifying reactions to local noise and reducing overall system stability.

\begin{center}
\fbox{
\parbox{0.94\linewidth}{
\textbf{Insights:}
LLM agents systematically replicate the human bias of underweighting the supply line, leading to myopic decisions. This finding is the primary explanation for why enhancing an isolated agent's cognitive capabilities can paradoxically worsen system instability, as superior intelligence amplifies, rather than corrects, myopic overreactions to local information.
}
}
\end{center}

\subsection{Do LLM Agents Exhibit Self-Interested Behavior?}

While myopia explains the agents' flawed system perception, it does not fully elucidate the motivational drivers behind their choices. To provide a comprehensive analysis, we identify a second, complementary mechanism: self-interested behavior, where agents prioritize local performance over global efficiency.

\begin{figure}[h]
\centering
\includegraphics[width=0.95\columnwidth]{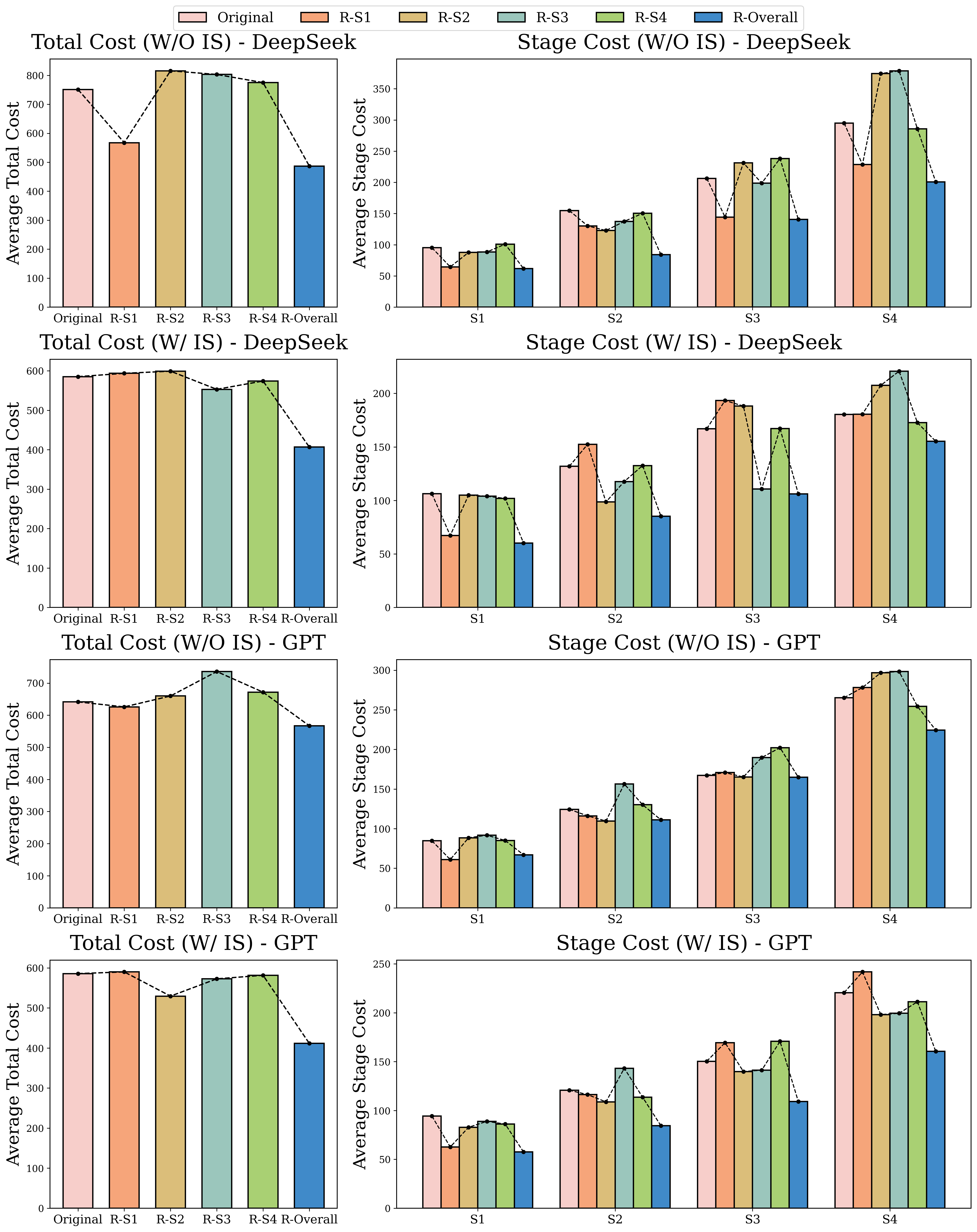} 
\caption{Total system cost and average stage cost.}
\label{fig:all_cost}
\end{figure}

\noindent\textbf{Experimental Results.}
This self-interested behavior is reflected in the system's cost structure (Figure~\ref{fig:all_cost})~\cite{akata2025playing}. Under information isolation, enhancing an upstream agent (S2, S3, or S4) reduces its local cost but increases the total cost of the supply chain beyond that of the original model. This reveals a fundamental misalignment between individual incentives and systemic performance metrics. Information sharing mitigates this misalignment. With shared information, selfish cost patterns disappear as agents' incentives align with global efficiency. The lowest total cost is achieved only when information sharing is combined with full agent enhancement (R-Overall).

\noindent\textbf{Discussion.} Our results show that under information isolation, LLM agents behave in a strongly self-interested manner~\cite{akata2025playing}. Enhanced agents optimize local costs based on limited information, but these locally rational actions increase system instability and harm collective performance. This outcome is driven by the interaction of self-interest and myopia. Myopia restricts the agent’s view of the system, while self-interest encourages actions that exploit this narrow perspective. Information sharing corrects both biases by providing global visibility and aligning individual decisions with system-wide efficiency.

\begin{center}
\fbox{
\parbox{0.94\linewidth}{
\textbf{Insights:}
LLM agents display self-interested and myopic behavior that prioritizes local gains over system performance, explaining why improving isolated agents can reduce efficiency. Information sharing mitigates this by aligning incentives with global objectives and enabling coordination.
}
}
\end{center}

\section{Conclusion}

This study presents a novel experimental paradigm integrating a Hierarchical Reasoning Framework with DeepSeek and GPT agents to analyze cognitive heterogeneity in a dynamic supply chain simulation. Results reveal three key characteristics: first, LLM agents demonstrate greater decision stability and statistical clarity than human benchmarks; second, they exhibit distinct myopic and self-interested behaviors, effectively replicating the cognitive biases that drive the bullwhip effect. We confirm that information sharing mitigates these latter tendencies. Furthermore, distinct behavioral signatures emerge: DeepSeek tends towards stability, whereas GPT displays greater volatility (see Appendix \ref{Appendix: DeepSeek and GPT}). Critically, these idiosyncratic traits diminish at the highest cognitive level, where both converge on a near-optimal strategy. This suggests rational task demands override model variance in highly capable agents, offering profound insights for AI-driven organizational research.

\section*{Limitations}

Despite validating the Hierarchical Reasoning Framework across distinct model families, our study has certain limitations. First, our reliance on advanced commercial models as proxies for cognitive depth means their underlying training data distributions and reinforcement learning objectives remain opaque. Consequently, we cannot perfectly isolate intrinsic strategic reasoning capabilities from potential training artifacts. Although our comparisons within the same model family strongly support the overall reasoning hierarchy, disentangling subtle stability differences linked to specific data mixtures remains an ongoing challenge. Second, while our setup embeds a deep agent within a shallow team, the framework does not explicitly model the adaptive capacity required to dynamically detect and respond to the bounded rationality of interactive counterparts. Finally, relying on the canonical Beer Distribution Game with deterministic parameters abstracts away the stochastic disruptions and complex network topologies of actual global supply chains. To establish broader generalizability, future research could extend this paradigm to other behavioral operations domains. Promising directions include exploring strategic adaptation based on regret in sealed bid auctions, performance variations under visible workloads in queueing systems, and profit margin biases within newsvendor inventory models.

\section*{Ethics Statement}

Our work utilizes LLMs to conduct traditional human behavior experiments, focusing on the analysis of heterogeneous cognitive abilities to understand LLM personality traits and their influence on overall experimental results. Through this data analysis, we aim to demonstrate the substitutability and usability of LLMs as experimental subjects. We hope to facilitate the adoption of LLMs for cost-effective and flexible large-scale experimentation. Overall, we anticipate no significant ethical or societal risks, as our primary objective is to investigate the controllability and validity of replicating canonical behavioral experiments using LLMs.

\section*{Acknowledgements}

This work was supported in part by the National Natural Science Foundation of China (No. 72293562, No. 72422016 and No. 724B2012).

\bibliography{sample}

\clearpage

\appendix

\section{The Beer Distribution Game: A Window into the Human Mind}
\label{Appendix: beer game}

\paragraph{The Mystery of the Bullwhip Effect}
Imagine a simple supply chain: a factory supplies a distributor, who supplies a wholesaler, who in turn supplies a retailer. One week, the retailer sees a tiny, almost unnoticeable uptick in customer beer sales. They order a little extra from the wholesaler to be safe. The wholesaler sees this slightly larger order and, fearing a trend, orders an even larger amount from the distributor. This chain reaction continues until it reaches the factory, where the initial tiny ripple in demand has become a tidal wave of panic-ordering. This phenomenon, where demand variability mysteriously amplifies as it moves up the supply chain, is known as the \textbf{bullwhip effect}. This phenomenon is not merely a statistical anomaly; rather, it represents a profoundly human narrative of decision-making under pressure and incomplete information. It is an emergent property of a complex, multi-agent strategic interaction conducted under conditions of pervasive uncertainty.

\paragraph{The Arena: The Beer Distribution Game}
To understand this mystery, researchers created a powerful simulation: the \textbf{Beer Distribution Game}. It’s a kind of "flight simulator" for managers, modeling a four-stage supply chain where the simple goal is to keep costs low by balancing two opposing evils: the cost of holding too much inventory and the cost of running out of stock (backorders). The genius of the game lies in its core constraint: \textbf{information asymmetry}. Like in the real world, you are trapped behind an "informational curtain." You can only see your own inventory and the orders coming from your immediate customer. You have no direct line to the true end-customer demand, nor can you see the inventory levels or decisions of your partners up the chain.

\begin{figure}[h]
  \centering
  \includegraphics[width=1\columnwidth]{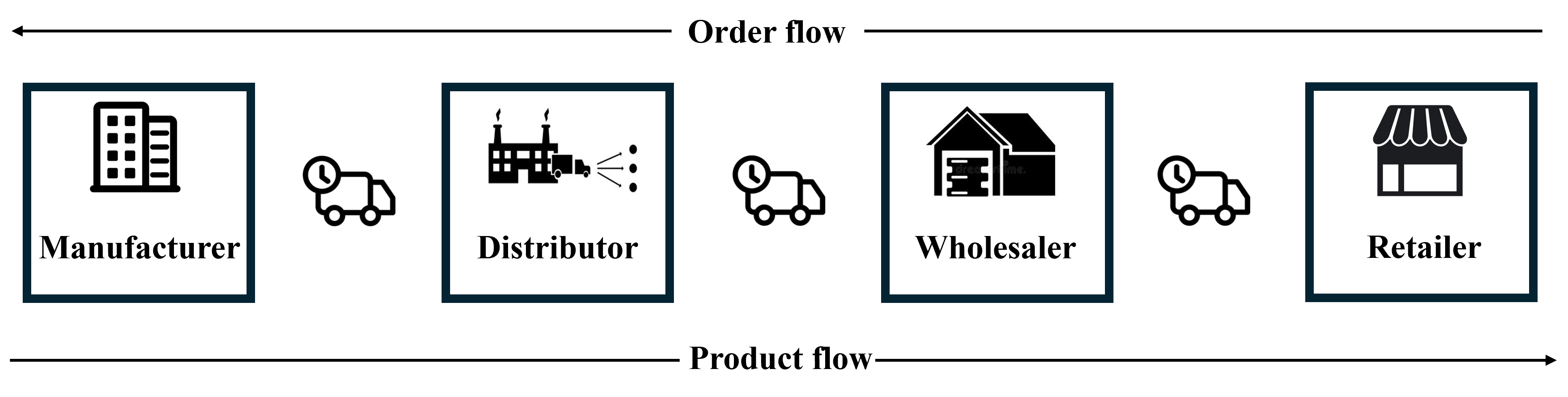}
  \caption{The Beer Distribution Game Structure.}
  \label{fig:beer_game_structure}
\end{figure}

\paragraph{From Logistics to Psychology}
At this point, the game ceases to be about simply managing stock; it transforms into a profound psychological and strategic puzzle. The challenge is no longer logistical but interpretive. To place an effective order, a player must essentially become a mind-reader, attempting to divine the intentions, fears, and strategies of the other invisible players from the sparest of clues. The single number they receive each week, the order quantity, is not clean data. It is a distorted signal, a faint echo of decisions made far downstream, with each intermediary adding their own layer of panicked over-correction or cautious undersupply. A truly strategic player understands this and must therefore deconstruct the signal, asking themselves:
\begin{center}
\textit{“Is this sudden surge in orders a real sign of a market boom? Or is it just my downstream partner panicking because they are about to stock out? How should I react, knowing my own decision will be misinterpreted and further amplified by the person upstream from me?”}
\end{center}
This constant, recursive second-guessing is the very heart of interdependent, strategic decision-making. It represents a crucial cognitive leap from simply reacting to numbers to reasoning about the minds behind them, forcing each player to build a mental model of their partners in the chain.

\paragraph{The Spectrum of Human Strategy}
For decades, experiments have consistently shown one thing: humans are not perfect optimizers. They panic, they hoard, they overreact, and the bullwhip effect reliably emerges. But simply calling this "bounded rationality" misses the most interesting part of the story: the sheer \textbf{heterogeneity} of human strategy. Observation reveals a spectrum of behaviors, ranging from simple, reactive decision-making on one end to more sophisticated attempts to forecast and anticipate system dynamics on the other. This variation in strategic depth is a critical driver of what happens to the system as a whole, yet it has been incredibly difficult to study systematically.

\paragraph{A New Lens: The Hierarchical Reasoning Framework}
Our work confronts this challenge by employing a \textbf{Hierarchical Reasoning Framework}. Instead of painting all players with the same brush, this framework provides a more intuitive and realistic way to model the observed spectrum of strategic thinking by organizing it into distinct cognitive tiers. For the context of this study, we focus on two fundamental levels:
\begin{itemize}[leftmargin=*]
    \item \textbf{Cognitively Shallow Thinkers:} These agents represent a baseline level of reasoning. They tend to be reactive, focusing primarily on their local state (their own inventory and the immediate orders they receive) and employing simpler heuristics to make decisions.
    \item \textbf{Cognitively Deep Thinkers:} These agents represent a more advanced level of reasoning. They possess the capacity for forward-looking, system-aware thinking, attempting to model the broader supply chain dynamics and anticipate how their actions might influence the system beyond their local position.
\end{itemize}
This layered model offers a powerful way to understand strategic interaction. The bullwhip effect, seen through this lens, can be conceptualized as the macroscopic result of the microscopic frictions and interactions between players operating at these different cognitive levels.

\paragraph{The Breakthrough: Large Language Models (LLMs) as Controllable Strategic Actors}
This reveals a fundamental challenge that has long constrained behavioral science: while we can observe heterogeneous strategies, we cannot reliably measure or control the internal cognitive state of a human participant. One cannot simply look inside a player's mind to verify their level of strategic reasoning. This measurement problem has made it nearly impossible to move beyond correlation and causally test how the cognitive \textit{composition} of a team truly affects its collective outcome.

Our research confronts this long-standing obstacle by pioneering the use of LLMs as a new class of experimental subject: \textit{controllable strategic actors}. Crucially, instead of trying to instruct a single model to "think harder" via prompts, we operationalize our Hierarchical Reasoning Framework by "casting" different, verifiably distinct models for each cognitive role. This allows us, for the first time, to construct virtual laboratories of economic interaction with a precisely specified and replicable mix of strategic abilities. We can now systematically assemble teams with one, two, or any number of "deep thinkers" and causally investigate how this cognitive heterogeneity impacts system-wide stability and efficiency, opening a new frontier for testing behavioral theories in multi-agent systems.

\section{Experimental Design: Crafting the Virtual Laboratory}
\label{Appendix: experimental design}

\paragraph{System Dynamics and Theoretical Framework}

To ensure our findings are grounded and comparable, we built our virtual laboratory on the bedrock of classical beer distribution game equations. This guarantees that our simulation faithfully mirrors the core dynamics of a four-stage, decentralized inventory system. All settings, variables, and rules are aligned with canonical models, allowing our results to speak directly to decades of prior research. Each team consists of four agents (S1-S4), and over 20 rounds, they make a single decision: how much to order. Their goal is to minimize their own local costs, which are a combination of holding too much stock or having too little. The precise mathematical rules governing shipments and inventory updates are as follows:

% \begin{align*}
% S_t^{i,g} &= \min\left\{ D_t, \max\left[ I_{t-1}^{i,g} + S_{t-2}^{i+1,g}, 0 \right] \right\} &&\text{for } i=1 \text{ (Retailer)}\\
%           &= \min\left\{ O_{t-2}^{i-1,g}, \max\left[ I_{t-1}^{i,g} + S_{t-2}^{i+1,g}, 0 \right] \right\} &&\text{for } i=2,3 \text{ (Wholesaler, Distributor)}\\
%           &= \min\left\{ O_{t-2}^{i-1,g}, \max\left[ I_{t-1}^{i,g} + O_{t-3}^{i,g}, 0 \right] \right\} &&\text{for } i=4 \text{ (Manufacturer)}
% \end{align*}
{\small % 或者用 \footnotesize 变得更小
\begin{align*}
S_t^{i,g} &= \min\left\{ D_t, \max\left[ I_{t-1}^{i,g} + S_{t-2}^{i+1,g}, 0 \right] \right\} 
\quad i=1 \\
&= \min\left\{ O_{t-2}^{i-1,g}, \max\left[ I_{t-1}^{i,g} + S_{t-2}^{i+1,g}, 0 \right] \right\}
\quad i=2,3 \\
&= \min\left\{ O_{t-2}^{i-1,g}, \max\left[ I_{t-1}^{i,g} + O_{t-3}^{i,g}, 0 \right] \right\}
\quad i=4
\end{align*}
}
The inventory at the end of each round is updated accordingly:
\begin{align*}
I_t^{i,g} &= I_{t-1}^{i,g} + S_{t-2}^{i+1,g} - S_t^{i,g} &&\text{for } i=1,2,3 \\
I_t^{4,g} &= I_{t-1}^{4,g} + O_{t-3}^{4,g} - S_t^{4,g}
\end{align*}

Here, the subscripts $t-2$ and $t-3$ explicitly denote the lead times: a 2-period delay for downstream shipments ($S_{t-2}^{i+1,g}$) and upstream orders ($O_{t-2}^{i-1,g}$), and a 3-period production delay for the manufacturer ($O_{t-3}^{4,g}$).

The cost function incentivizes each agent to balance their inventory:
% \begin{equation*}
% C_i(T) = \sum_{t=1}^{T} \left[ h \cdot \max\left\{ I_t^{i,g}, 0 \right\} - s \cdot \min\left\{ I_t^{i,g}, 0 \right\} \right]
% \end{equation*}
\begin{equation*}
C_i(T) = \sum_{t=1}^{T} \left[ h^i \max\left\{ I_t^{i,g}, 0 \right\} - s^i \min\left\{ I_t^{i,g}, 0 \right\} \right]
\end{equation*}
All core simulation parameters were held constant to isolate the effects of agent cognition and information structure.

\begin{table}[h]
\centering
\renewcommand{\arraystretch}{1.3}
\begin{tabularx}{\linewidth}{lX}
\toprule
\textbf{Parameter} & \textbf{Value / Description} \\
\midrule
Simulation Horizon ($T$) & 20 periods. \\
Replications & 32 independent runs for each experimental configuration for statistical robustness. \\
Customer Demand ($D_t$) & Drawn from a stationary uniform distribution, $U[0, 8]$. Only the retailer (S1) observes the demand draws directly. \\
Lead Times & 2 periods for retailer, wholesaler, and distributor; 3 periods for the manufacturer. \\
Initial Inventory & 12 units for each stage. \\
% stage Capacity & 20 units per period (maximum production/shipment). \\
Unit Holding Cost ($h$) & 0.5 per unit per period. \\
Unit Backlog Cost ($s$) & 1.0 per unit per period. \\
\bottomrule
\end{tabularx}
\caption{Core Parameters of the Simulation Environment}
\label{tab:exp_params_detailed}
\end{table}

\paragraph{Hypotheses}
Our study is structured around three central hypotheses that challenge the rationality of decision-making in this environment:
\begin{description}[style=unboxed, leftmargin=0cm]
  \item[\textbf{(H1)}] \emph{The bullwhip effect does not occur when the demand distribution is known and stationary.}
  \item[\textbf{(H2)}] \emph{Sharing dynamic inventory information (IS) across the supply chain reduces the level of order oscillation.}
  \item[\textbf{(H3)}] \emph{Participants will not underweight the pipeline supply when demand is known and stationary, regardless of IS.}
\end{description} 

\paragraph{Agent Implementation and Rationale}
Our research is designed to empirically investigate a Hierarchical Reasoning Framework, which posits that strategic thinking occurs in distinct cognitive layers of increasing sophistication. A primary challenge in testing this framework is isolating the variable of cognitive depth from other confounding factors. To overcome this, we developed a novel, dual-family methodology to operationalize these cognitive layers. We established a \textbf{cognitively shallow tier} using \textbf{DeepSeek-V3 and GPT-4.1} to represent baseline decision-making, and a \textbf{cognitively deep tier} with \textbf{DeepSeek-R1 and the cutting-edge GPT-5} to represent advanced, system-aware reasoning.

This experimental design allows for a robust test of the framework's core tenets. The DeepSeek series, with its shared architectural lineage, provides a controlled setting to isolate the effects of advancing up the cognitive hierarchy. To ensure that our conclusions about the framework are generalizable and not an artifact of a single model family, we benchmark our findings against the state-of-the-art GPT series, with the novel inclusion of GPT-5 providing a particularly stringent test. The reliability of this setup is further reinforced by evidence that experimental outcomes are robust to temperature variations across these models, providing a compelling foundation for our analysis of hierarchical reasoning. To ensure all agents produce a transparent and structured reasoning process, they were guided by Chain-of-Thought (CoT) prompting.

\paragraph{Experimental Conditions}
To test our hypotheses using the agents defined above, we designed a suite of experiments built upon our Hierarchical Reasoning Framework. All conditions were replicated across both the DeepSeek and GPT model families to ensure robust and generalizable findings. The experiments feature two primary conditions:

\begin{itemize}[leftmargin=*]
    \item \textbf{Homogeneous Conditions:} These are our control groups, establishing baselines for uniformly skilled teams.
    \begin{itemize}
        \item \emph{Original:} A team of four cognitively shallow agents.
        \item \emph{R-Overall:} A team of four cognitively deep agents.
    \end{itemize}

    \item \textbf{Hierarchical Conditions:} These are our primary experimental groups, creating "mixed-ability" teams to test the impact of heterogeneity. We place a single deep agent in an otherwise shallow team, rotating its position to see where intelligence matters most.
    \begin{itemize}
        \item \emph{R-S1:} The deep agent is the Retailer.
        \item \emph{R-S2:} The deep agent is the Wholesaler.
        \item \emph{R-S3:} The deep agent is the Distributor.
        \item \emph{R-S4:} The deep agent is the Manufacturer.
    \end{itemize}
\end{itemize}

\section{Supplementary Evidence: Why GPT-5 and DeepSeek-R1 Are Classified as Deep Models}
\label{Appendix: cognitive choice}

We classify GPT-5 and DeepSeek-R1 as deep models based on two complementary considerations: (\emph{i}) their architectural design and benchmark performance, and (\emph{ii}) their observed behavioral performance in our simulation environment.

First, from an architectural and benchmark-based perspective, DeepSeek-R1 and GPT-5 are explicitly designed as reasoning-oriented models with endogenous multi-step reasoning capabilities. By contrast, DeepSeek-V3 and GPT-4.1 primarily function as base models without intrinsic reasoning modules. This architectural distinction is also reflected in benchmark performance. Across widely adopted reasoning-intensive benchmarks, including \textit{AIME}, \textit{GPQA}, \textit{MMLU}, and \textit{SWE-bench}, DeepSeek-R1 and GPT-5 consistently outperform their corresponding base counterparts. These benchmarks evaluate mathematical reasoning, graduate-level problem solving, and software engineering reasoning capabilities, which are closely aligned with structured decision-making tasks. Hence, benchmark evidence provides an important empirical basis for our classification.

Second, our classification is further supported by simulation-based behavioral evidence. Rather than relying solely on benchmark claims, we examine whether the distinction between reasoning-oriented and base models produces systematically different outcomes in a controlled dynamic supply chain simulation. The results show that, across both model families, the reasoning-oriented configuration consistently achieves lower \emph{total cost} and lower \emph{order variance} than the base configuration.

\begin{table}[h]
\centering
\setlength{\tabcolsep}{1pt}
\resizebox{\columnwidth}{!}{%
\begin{tabular}{llrrrr}
\toprule
\multicolumn{6}{c}{\textbf{Model A: DeepSeek}} \\
\midrule
Config. & Cond. & Total Cost & Var (Mean) & Var (Med) & Var (Std) \\
\midrule
R-Overall & W/O IS & 486.95 & 29.43 & 12.29 & 63.54 \\
R-Overall & W/ IS  & 406.80 & 17.71 & 9.47  & 26.99 \\
Original  & W/O IS & 751.27 & 39.43 & 18.73 & 65.14 \\
Original  & W/ IS  & 585.34 & 20.15 & 11.12 & 29.88 \\
\midrule
\multicolumn{6}{c}{\textbf{Model B: GPT}} \\
\midrule
Config. & Cond. & Total Cost & Var (Mean) & Var (Med) & Var (Std) \\
\midrule
R-Overall & W/O IS & 567.10 & 31.42 & 13.88 & 46.09 \\
R-Overall & W/ IS  & 411.54 & 13.72 & 11.88 & 9.09  \\
Original  & W/O IS & 641.55 & 58.45 & 23.07 & 89.86 \\
Original  & W/ IS  & 585.88 & 40.39 & 20.70 & 71.44 \\
\bottomrule

\end{tabular}%
}
\caption{Supplementary Evidence Supporting the Classification of GPT-5 and DeepSeek-R1 as Deep Models. ``Config.'' denotes configuration, and ``Cond.'' denotes experimental condition.}
\label{tab:deep_model_behavior}
\end{table}

The magnitude and consistency of these differences provide clear behavioral validation for distinguishing reasoning-oriented models from base models in dynamic multi-agent settings.

\section{The Prompt: Scripting the Agent's Worldview}
\label{Appendix: prompt section}

To bring our experimental design to life, we developed a meticulous prompting strategy that serves as the "script" and "director's cues" for our LLM agents. Our core methodological principle is one of \textit{casting, not coaching}. We do not try to coax a single model into acting "smarter" or "more strategic" through complex instructions. Instead, as previously defined, we cast verifiably distinct models for the cognitively shallow and deep roles. This ensures that the behavioral differences we observe stem from the models' genuine, intrinsic capabilities, not their ability to follow creative writing instructions.

To maintain this focus on intrinsic abilities, our prompts were designed with scientific minimalism in mind. We deliberately avoided any "method acting" instructions or persona-based cues (like "you are a risk-averse manager"). The script is purely informational and objective: it provides the universal rules of the game, the agent's current situation, and its unwavering goal to minimize costs. This rigorous approach ensures we are studying the models' fundamental reasoning patterns, not artifacts of the prompt design. The only variable manipulated within the script itself is the agent's access to information.

\paragraph{Information Structure System Messages}
At the very start of each simulation, a one-time system message establishes the fundamental "rules of the universe" for the agent, defining its worldview for the entire 20-round game. This initial prompt places the agent into one of two starkly different informational realities.

In the classic \textbf{isolated information} setting, the message draws the "informational curtain," plunging the agent into a state of local perception and forcing it to navigate the fog of uncertainty.

\begin{tcolorbox}[colback=white, colframe=black!50, sharp corners=south, breakable, title=System Prompt: Isolated Information (Baseline)]
You have \textbf{NO VISIBILITY} into the inventory or backlog levels at other stages of the supply chain. You must make decisions \textbf{solely based on your own local state} (demand, inventory, backlog, and arriving deliveries).
\end{tcolorbox}

In the contrasting \textbf{information sharing} condition, the curtain is lifted. The message grants the agent a "god's-eye view" of the entire system, creating the potential for globally coordinated, omniscient decisions.

\begin{tcolorbox}[colback=white, colframe=black!50, sharp corners=south, breakable, title=System Prompt: Information Sharing]
\textbf{IMPORTANT NOTE:} You have \textbf{FULL VISIBILITY} of the \textbf{inventory} and \textbf{backlog} levels across \textbf{ALL stages} -- retailer, wholesaler, distributor, and manufacturer. Please use this shared information to \textbf{make globally optimal decisions}.
\end{tcolorbox}

\paragraph{Dynamic Process Messages}
With the worldview established, the game progresses through dynamic, turn-by-turn "director's cues" in the form of process messages. This message is the complete and sole prompt for each round, containing all the information the agent is permitted to see before making its decision. The content of this cue differs dramatically depending on which of the two worlds the agent inhabits.

In the isolated world, the message is sparse and local, reinforcing the agent's limited perspective.

\begin{tcolorbox}[colback=white, colframe=black!50, sharp corners=south, breakable, title=Process Message: Isolated Information]
Now this is round \texttt{\{period + 1\}}, and you are at stage \texttt{\{stage + 1\}} of \texttt{\{len(state\_dict)\}} in the supply chain.

You have \textbf{no visibility} into the inventory or backlog levels at other stages. Please make your decision \textbf{solely based on your own current state and local information}.

\textit{(Retailer only):} The demand at the retailer (stage 1) is \texttt{\{demand1\}}.

Given your current state:
\texttt{\{get\_state\_description(stage\_state)\}}

\textit{(Non-retailer stages):} Your downstream order from stage \texttt{\{stage\}} for this round is \texttt{\{action\_dict.get(f'stage\_\{stage - 1\}', 'N/A')\}}.

What is your action (order quantity) for this round? Your aim is to minimize the cost, where one unit of backlog costs 1.0 and one unit of inventory costs 0.5. Please provide your action as a non-negative integer within brackets at the end of your response (e.g., [0]).
\end{tcolorbox}

In the transparent world of information sharing, the message is enriched with a complete, real-time data feed from across the entire supply chain, empowering the agent with global awareness.

\begin{tcolorbox}[colback=white, colframe=black!50, sharp corners=south, breakable, title=Process Message: Information Sharing]
Now this is round \texttt{\{period + 1\}}, and you are at stage \texttt{\{stage + 1\}} of \texttt{\{len(state\_dict)\}} in the supply chain.

\textbf{IMPORTANT NOTE:} You have \textbf{FULL VISIBILITY} of the inventory and backlog levels across \textbf{ALL stages}. Please \textbf{thoroughly review} this information before making your decision.

The inventory levels of retailer, wholesaler, distributor, and manufacturer are \texttt{\{state\_dict['stage\_0']['inventory']\}}, \texttt{\{state\_dict['stage\_1']['inventory']\}}, \texttt{\{state\_dict['stage\_2']['inventory']\}}, and \\
\texttt{\{state\_dict['stage\_3']['inventory']\}} respectively.

The current backlog levels of retailer, wholesaler, distributor, and manufacturer are: \texttt{\{state\_dict['stage\_0']['backlog']\}}, \texttt{\{state\_dict['stage\_1']['backlog']\}}, \texttt{\{state\_dict['stage\_2']['backlog']\}}, and\\
\texttt{\{state\_dict['stage\_3']['backlog']\}} respectively.

\textit{(Retailer only):} The demand at the retailer (stage 1) is \texttt{\{demand1\}}.

Given your current state:
\texttt{\{get\_state\_description(stage\_state)\}}

\textit{(Non-retailer stages):} Your downstream order from stage \texttt{\{stage\}} for this round is \texttt{\{action\_dict.get(f'stage\_\{stage - 1\}', 'N/A')\}}.

What is your action (order quantity) for this round? Your aim is to minimize the cost, where one unit of backlog costs 1.0 and one unit of inventory costs 0.5. Please provide your action as a non-negative integer within brackets at the end of your response (e.g., [0]).
\end{tcolorbox}

\section{Supplementary Outcome: Regression Analysis Details}
\label{Appendix: regression}

Table~\ref{tab:pvalues_regression} reports the $p$-values for the regression analysis assessing supply line underweighting in LLM agents. The regression follows the form described in Equation~\ref{eq1}, with the key result being that $a_N > a_I$ in all tested conditions. The sign test confirms this myopic underweighting effect is statistically significant across all models and replications.

Moreover, under the Tracking Demand Policy, order quantities do not exhibit statistically significant myopic behavior ($p = 0.99$). By contrast, both IPPO and MAPPO display statistically significant myopic behavior, with IPPO showing strong significance ($p < 0.001$) and MAPPO also remaining significant ($p = 0.021$). These findings are consistent with the preceding results and further underscore the advantages of LLM-based simulation over heuristic algorithms and traditional reinforcement learning based approaches.

\begin{table}[h]
\centering
\small
\begin{tabular}{lcccc}
\toprule
 & \multicolumn{2}{c}{\textbf{W/O IS}} & \multicolumn{2}{c}{\textbf{W/ IS}} \\
\cmidrule(lr){2-3} \cmidrule(lr){4-5}
\textbf{Case} & \textbf{DS} & \textbf{GPT} & \textbf{DS} & \textbf{GPT} \\
\midrule
Original    & $<0.001$ & $<0.001$ & $<0.001$ & $<0.001$ \\
R-S1        & $<0.001$ & $<0.001$ & $<0.001$ & $<0.001$ \\
R-S2        & $<0.001$ & $<0.001$ & $<0.001$ & $<0.001$ \\
R-S3        & $<0.001$ & $<0.001$ & $<0.001$ & $<0.001$ \\
R-S4        & $<0.001$ & $<0.001$ & $<0.001$ & $<0.001$ \\
R-Overall   & $<0.001$ & $<0.001$ & $<0.001$ & $<0.001$ \\
\bottomrule
\end{tabular}
\caption{$p$-values for supply line underweighting regression.}
\label{tab:pvalues_regression}
\end{table}

\section{Supplementary Analysis: Do DeepSeek and GPT Have Different Personalities in Games?}
\label{Appendix: DeepSeek and GPT}

Beyond the general behaviors observed, a key question is whether different LLM families exhibit distinct and stable personalities in strategic play. To investigate this question systematically, we conducted a direct and statistically rigorous comparison between the DeepSeek and GPT series using a MWU test on their performance outcomes (Table~\ref{tab:your_label}). The results reveal a compelling and informative dichotomy. In 9 out of 12 conditions, a significant difference emerges, with \textbf{DeepSeek consistently exhibiting markedly lower variance}. This suggests that the two model families possess clear and reproducible behavioral signatures in this dynamic multi-agent environment: \textbf{DeepSeek trends towards more stable and conservative decision-making, whereas GPT displays greater behavioral variability and volatility}.

% \begin{table*}[t]
% \centering
% \setlength{\tabcolsep}{2.8pt}
% \caption{One-tail Mann–Whitney U test $p$-values comparing DeepSeek and GPT's outcome.}
% \begin{tabular}{lcccccc}
% \toprule
% \textbf{Con} & \textbf{R-Original} & \textbf{R-Overall} & \textbf{R-S1} & \textbf{R-S2} & \textbf{R-S3} & \textbf{R-S4} \\
% \midrule
% \textbf{W/O IS}  & 0.009** &  0.247 &   $<$0.001*** &   0.018* & 0.012* &   0.239  \\
% \textbf{W/ IS}  &   $<$0.001*** &   0.372   &   $<$0.001*** &   0.001** &    0.046* &    $<$0.001*** \\
% \bottomrule
% \end{tabular}
% \label{tab:your_label}
% \end{table*}

\begin{table}[t]
\centering

\begin{tabular}{lcc}
\toprule
\textbf{Case} & \textbf{W/O IS} & \textbf{W/ IS} \\
\midrule
R-Original & 0.009 & $<$0.001 \\
R-Overall  & 0.247        & 0.372       \\
R-S1       & $<$0.001 & $<$0.001 \\
R-S2       & 0.018 & 0.001 \\
R-S3       & 0.012 & 0.046 \\
R-S4       & 0.239        & $<$0.001 \\
\bottomrule
\end{tabular}
\caption{One-tail Mann–Whitney U test $p$-values comparing DeepSeek and GPT's outcome.}
\label{tab:your_label}
\end{table}

However, the most intriguing finding arises from the R-Overall condition, where the most capable agents are deployed. Here, we observe no statistically significant difference in performance between the two model series, regardless of information sharing. This counter-intuitive result, consistent with our earlier cost and variance analyses (Figure~\ref{fig:stage_order} and \ref{fig:all_cost}), offers a powerful insight: while the models may have markedly different inherent personalities at lower levels of sophistication, \textbf{these stylistic differences dissolve at the highest level of cognitive capability}. Instead of their inherent biases dominating, both DeepSeek and GPT independently converge on a shared, near-optimal strategy dictated by the rational logic of the supply chain itself. This implies that for complex, system-level tasks, the choice between top-tier models may become less critical, as both ultimately identify a similar path to effective performance.

\end{document}